\documentclass{aa}
\usepackage{graphicx}
\usepackage{natbib}
\newcommand{\hei}{He~I}
\newcommand{\sii}{Si~I}

\usepackage{color}
\begin{document}
\headnote{}
   \title{Comparing magnetic field extrapolations with measurements of magnetic loops.}


   \author{T. Wiegelmann, A. Lagg, S.K. Solanki,  B. Inhester,  J. Woch
          }

   \institute{Max-Planck-Institut f\"ur Sonnensystemforschung,
Max-Planck-Stra\ss{}e 2, 37191 Katlenburg-Lindau, Germany}
     \offprints{T. Wiegelmann,
\email{wiegelmann@linmpi.mpg.de}  }
%
\date{Received 6 September 2004 / Accepted 2 December 2004 \\
DOI: 10.1051/0004-6361:20042421 \\
Publication: Astronomy and Astrophysics, Volume 433, Issue 2, April II 2005, pp.701-705
}
\abstract{We compare magnetic field extrapolations from a photospheric
magnetogram with the observationally inferred structure of magnetic loops in a
newly developed active region. This is the first time that the reconstructed
3D-topology of the magnetic field is available to test the extrapolations.  We
compare the observations with potential fields, linear force-free fields and
non-linear force-free fields.  This comparison reveals that a potential field
extrapolation is not suitable for a reconstruction of the magnetic field in
this young, developing active region.  The inclusion of field-line-parallel
electric currents, the so called force-free approach, gives much better
results. Furthermore, a non-linear force-free computation reproduces the
observations better than the linear force-free approximation, although no free
parameters are available in the former case.

\keywords{Sun: magnetic fields -Sun: corona -Sun: chromosphere} }
\titlerunning{Loops} \authorrunning{Wiegelmann et al.}

   \maketitle
%
\section{Introduction}
Due to the low plasma $\beta$ the magnetic field is the dominating quantity in
the low solar corona. Thus, the 3-D magnetic field structure is of basic
importance for
physical processes in the solar atmosphere, such as flares, coronal mass
ejections and X-ray jets.  Direct observations of chromospheric and coronal
magnetic fields are difficult, but significant progress has been made within
the last few years, e.g.
\cite{lee99,lin00,kundu01,white02,raouafi02,solanki03,lagg04,lin04}.
Here we compare magnetic loops reconstructed from magnetic field measurements
by \cite{solanki03} with magnetic fields extrapolated from a photospheric
magnetogram. The measured fields allow a much more sensitive test of
extrapolations than other observations.
\section{Measurements of magnetic fields in the upper chromosphere}
\label{measure}
The inference of the magnetic vector is based on an inversion
technique applied to spectropolarimetric data of the photospheric \sii{} line
at 1082.7~nm and the chromospheric \hei{} 1083~nm triplet. The data were
recorded with the Tenerife Infrared Polarimeter mounted on the German Vacuum
Tower Telescope (VTT). The spatial resolution
of the data was limited by seeing to 1.5\arcsec{}.

The photospheric magnetic vector map was obtained by applying the inversion
code SPINOR to the \sii{} Stokes profiles \cite[]{frutiger:00a}. The \hei{}
triplet provided the maps of the chromospheric vector magnetic field. This
triplet, which has a complex non-LTE line formation \cite[]{avrett:94} but is
nearly optically thin \cite[]{giovanelli:77}, was analysed by applying the
Unno-Rachkowsky  solution \cite[]{unno56,rachkowsky:67}
to describe the individual Zeeman components of each member
of the triplet, together with a simple
implementation of the Hanle effect based on recent developments.
These have
convincingly demonstrated that the Hanle effect in forward scattering close to
the solar disk center creates measurable linear polarization in spectral lines
\cite[]{trujillobueno:02}.

The inclination
and azimuthal angle from the chromospheric magnetic field map was used to
trace magnetic field lines.  We identified field lines as magnetic loops if
the following criteria were fulfilled: the magnetic field strength must
decrease with height, the inclination and azimuthal angles must not vary
strongly from one pixel to the other and the height of the two footpoints must
be similar.  For a more detailed description of the observations and the
analysis technique we refer to \cite{lagg04} and \cite{solanki03}.

We are aware of the assumptions and approximations which enter the
magnetic field reconstruction from the polarimetric observations: The
Milne-Eddington approach neglects vertical gradients and the simple
implementation of the Hanle-effect restricts the reliable determination of the
azimuthal angle to regions where the magnetic field is strongly inclined
to the line of sight. Furthermore this method neglects the changes in the
polarization signal caused by the incomplete Paschen-Back splitting
\cite[]{socas-navarro04}.
From a preliminary comparison of the obtained
H$\alpha$ slit jaw images and the inferred magnetic loops
we are confident that the
retrieved magnetic field topology is close to the real situation.
In the following we use the term "observed loops" to name the loops
inferred under these assumptions and to distinguish them from loops
computed with the help of extrapolations from a photospheric magnetogram.
%
\section{Computation of 3D magnetic fields from photospheric magnetic field measurements.}
\label{extrapol}
A number of authors have modelled the coronal magnetic field by extrapolating
from photospheric magnetic field observations. It is generally assumed that
the magnetic pressure in the corona is much higher than the plasma pressure
(small plasma $\beta$) and that therefore the magnetic field is nearly
force-free. The extrapolation methods based on this assumption include
potential field extrapolation (e.g. \citeauthor{semel67}, \citeyear{semel67}),
linear force-free field extrapolation
(e.g. \citeauthor{chiu77}, \citeyear{chiu77};
\citeauthor{seehafer78}, \citeyear{seehafer78};
\citeauthor{seehafer82}, \citeyear{seehafer82};
\citeauthor{semel88}, \citeyear{semel88})
and nonlinear force-free field extrapolation
(e.g. \citeauthor{sakurai81}, \citeyear{sakurai81};
\citeauthor{roumeliotis96}, \citeyear{roumeliotis96};
\citeauthor{amari97}, \citeyear{amari97};
\citeauthor{amari99}, \citeyear{amari99};
\citeauthor{wheatland00}, \citeyear{wheatland00};
\citeauthor{tw04}, \citeyear{tw04}).
Force-free magnetic fields have to obey the equations
\begin{eqnarray}
(\nabla \times {\bf B }) \times{\bf B} & = & {\bf 0},
\label{forcefree} \\
\nabla \cdot{\bf B}    & = &         0.
\label{solenoidal-ff}
\end{eqnarray}
which are equivalent to
\begin{eqnarray}
(\nabla \times {\bf B }) & = & \alpha {\bf B}, \\
{\bf B } \cdot \nabla \alpha & = & 0.
\end{eqnarray}
In general $\alpha$ is a function of space.
Taking this into account corresponds to the  non-linear
force-free approach. A popular simplification is to choose $\alpha={\rm
constant}$ in the entire computational domain, the linear force-free
approach. The choice
$\alpha=0$ corresponds to current-free potential fields.  In this paper we
compute potential fields, linear force-free and non-linear force-free fields
and compare the result with the magnetic loops reconstructed from the
observations.
\subsection{Potential and linear force-free fields.}
We use the method of \cite{seehafer78} for calculating the linear force-free
field.  The method requires a line-of-sight magnetogram and contains a free
scalar parameter $\alpha$, where $\alpha=0$ corresponds to potential fields.
The method gives the components of the magnetic field in terms of a Fourier
series.  The observed magnetogram which covers a rectangular region extending
from $0$ to $L_x$ in $x$ and $0$ to $L_y$ in $y$ is artificially extended onto
a rectangular region covering $-L_x$ to $L_x$ and $-L_y$ to $L_y$ by taking an
antisymmetric mirror image of the original magnetogram in the extended region,
i.e.  $B_z(-x,y) = -B_z(x,y)$ and $B_z(x,-y) = -B_z(x,y)$.  The advantage of
taking the antisymmetric extension of the original magnetogram is that the
extended magnetogram is automatically flux balanced.  We use a Fast Fourier
Transformation (FFT) scheme to determine the coefficients of the Fourier
series. $\alpha$ has the dimension of an inverse length. As a characteristic
length scale we choose the harmonic mean $L$ of $L_x$ and
$L_y$.
(See \citeauthor{seehafer78} \citeyear{seehafer78} for details.)
\subsection{Non-linear force-free fields.}
We solve  Eqs. (\ref{forcefree}) and (\ref{solenoidal-ff}) by means
of an optimization principle (\citeauthor{wheatland00} \citeyear{wheatland00},
 \citeauthor{tw04} \citeyear{tw04}):
\begin{equation}
L=\int_{V}  w(x,y,z) \, \left[B^{-2} \, |(\nabla \times {\bf B}) \times {\bf
B}|^2 +|\nabla \cdot {\bf B}|^2\right] \, d^3x
\label{defL1},
\end{equation}
where $w(x,y,z)$ is a weighting function.  It is obvious that (for $w>0$) the
force-free Eqs. (\ref{forcefree}-\ref{solenoidal-ff}) are fulfilled when
L equals zero.  We compute the magnetic field in a box with $nx=95$, $ny=68$
and $nz=40$ points.  The numerical method works as follows.  As an initial
configuration we compute a potential magnetic field in the whole box with the
help of the \cite{seehafer78} method.  As the next step we use photospheric
vector magnetic field data to prescribe the bottom boundary (photosphere) of the
computational box. On the lateral and top boundaries the field is chosen from
the potential field above.  We iterate for the magnetic field inside the
computational box by minimizing Eq. (\ref{defL1}). The weighting function $w$
equals 1 everywhere in the computational box except in a boundary layer of $10$
points towards the lateral and top boundary of the computational box, where
$w$ decreases smoothly to 0 with a cosine function.
The boundary layer diminishes the influence
of the lateral and top boundary conditions onto the magnetic field in the box.
(See \citeauthor{tw04} \citeyear{tw04} for details.)
\section{Results}
\label{compare}
\begin{figure*}
\centering
\mbox{
\includegraphics[bb=40 15 380 195,clip,height=4cm]{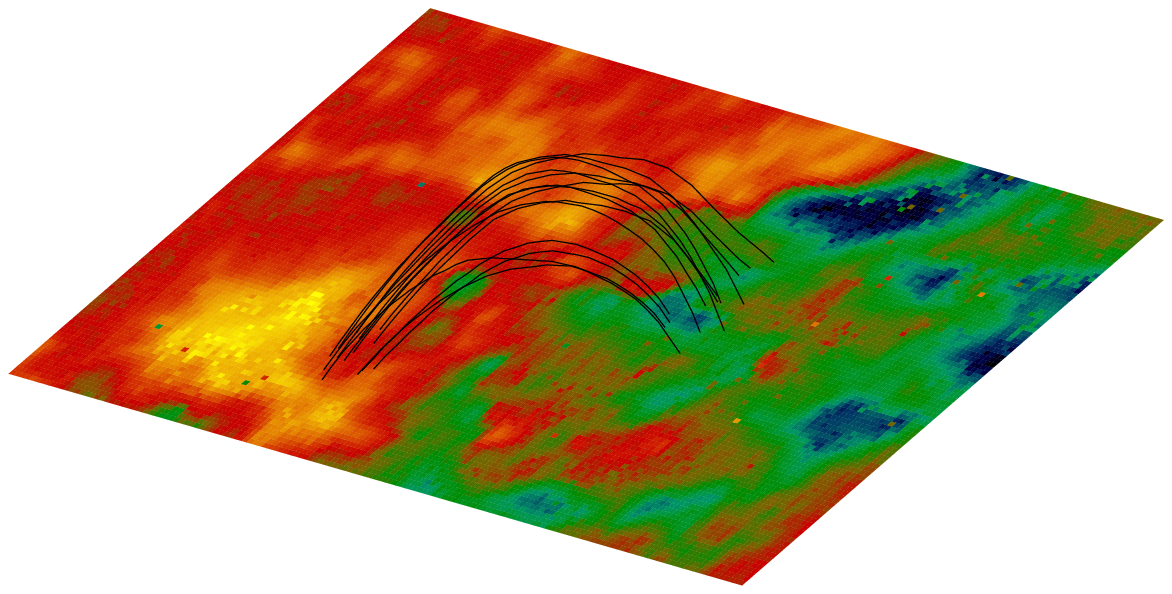}
\includegraphics[bb=120 100 270 290,clip,height=4cm,width=7cm]{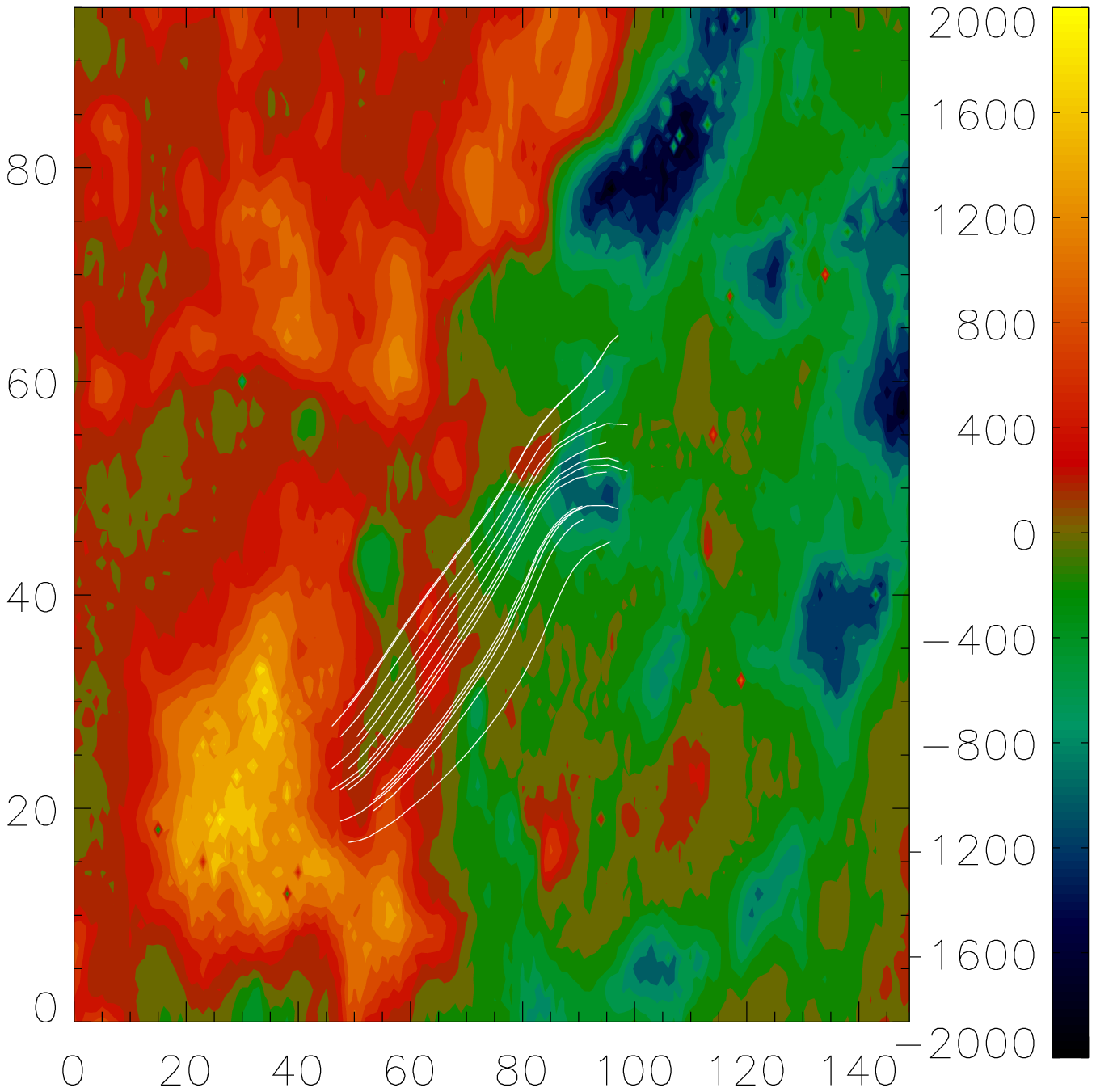}
}
\mbox{
\includegraphics[bb=40 15 380 195,clip,height=4cm]{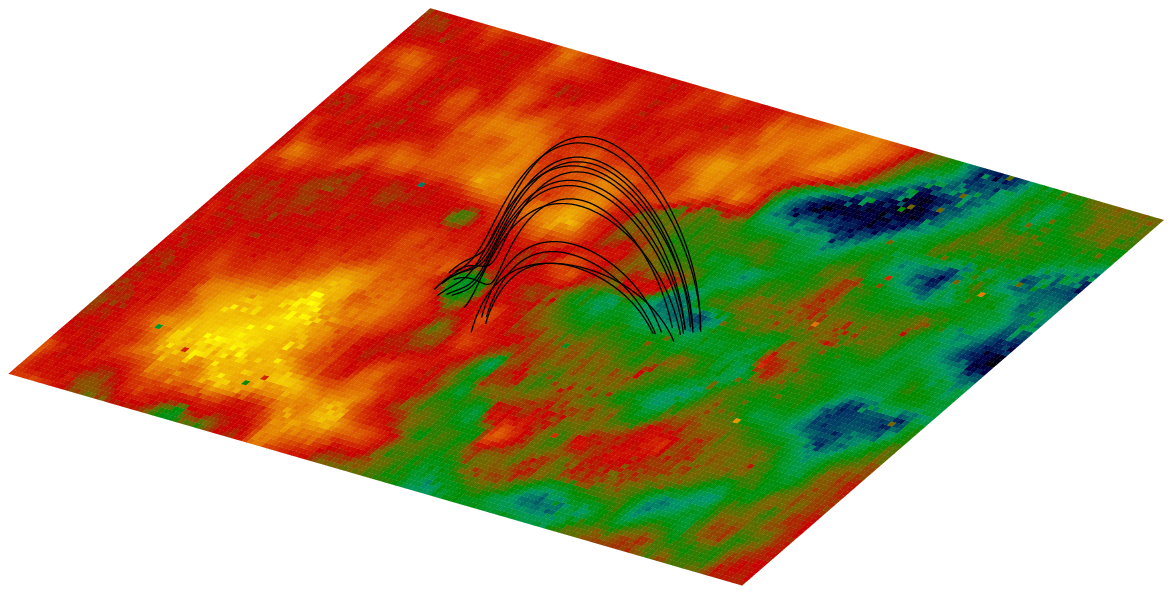}
\includegraphics[bb=120 100 270 290,clip,height=4cm,width=7cm]{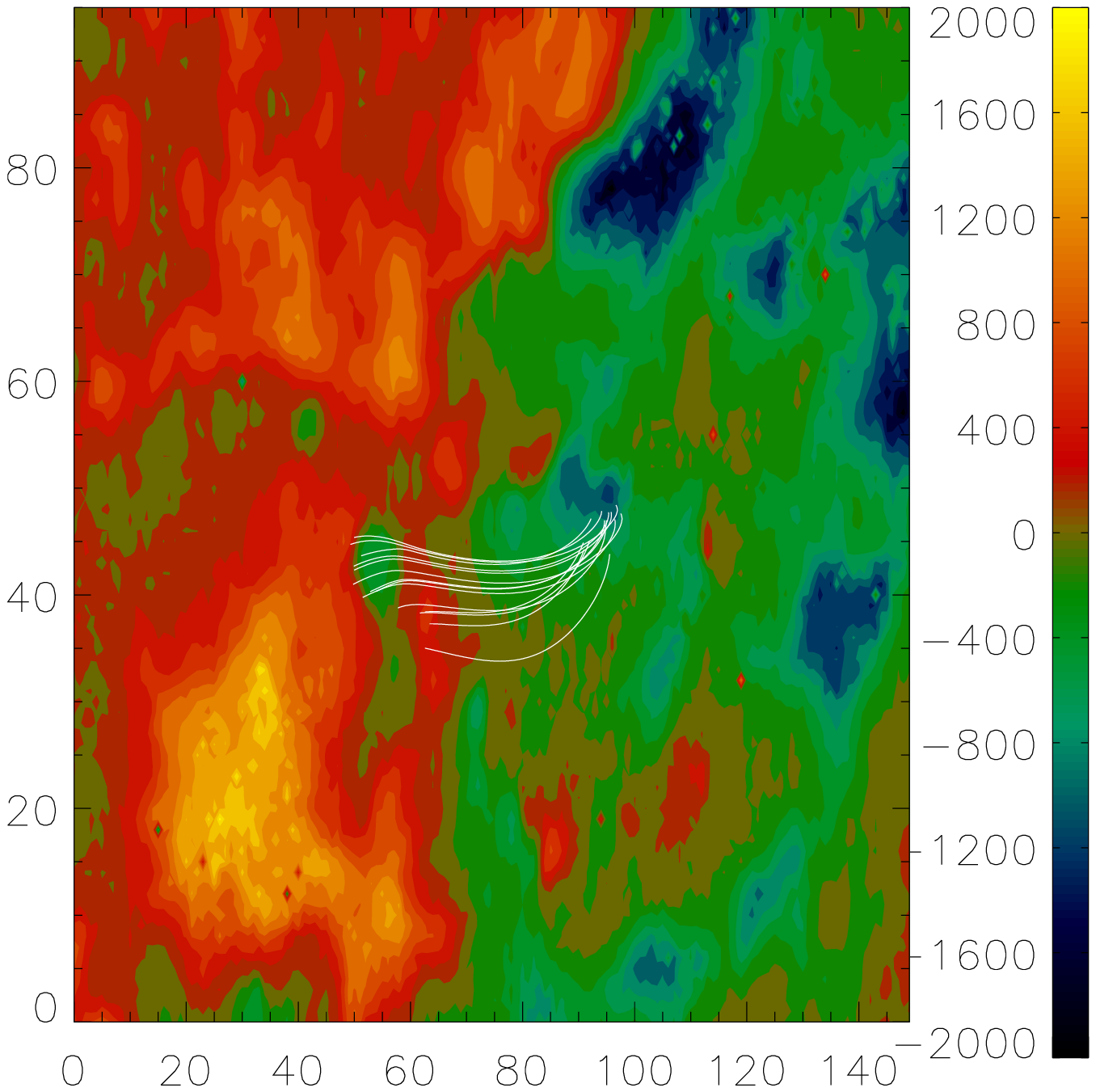}
}
\mbox{
\includegraphics[bb=40 15 380 195,clip,height=4cm]{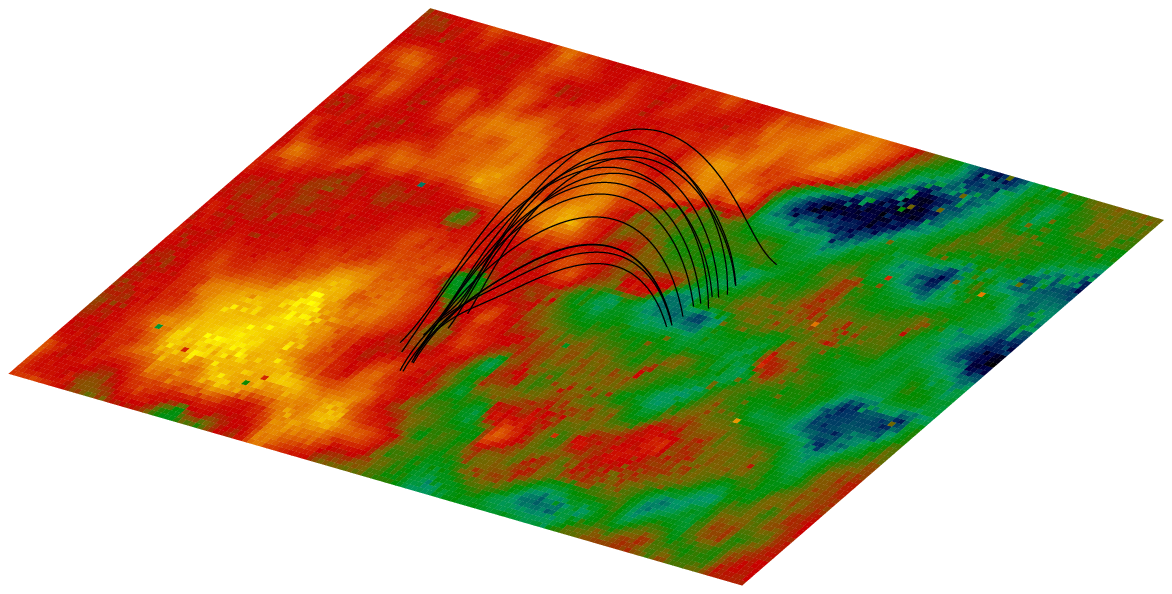}
\includegraphics[bb=120 100 270 290,clip,height=4cm,width=7cm]{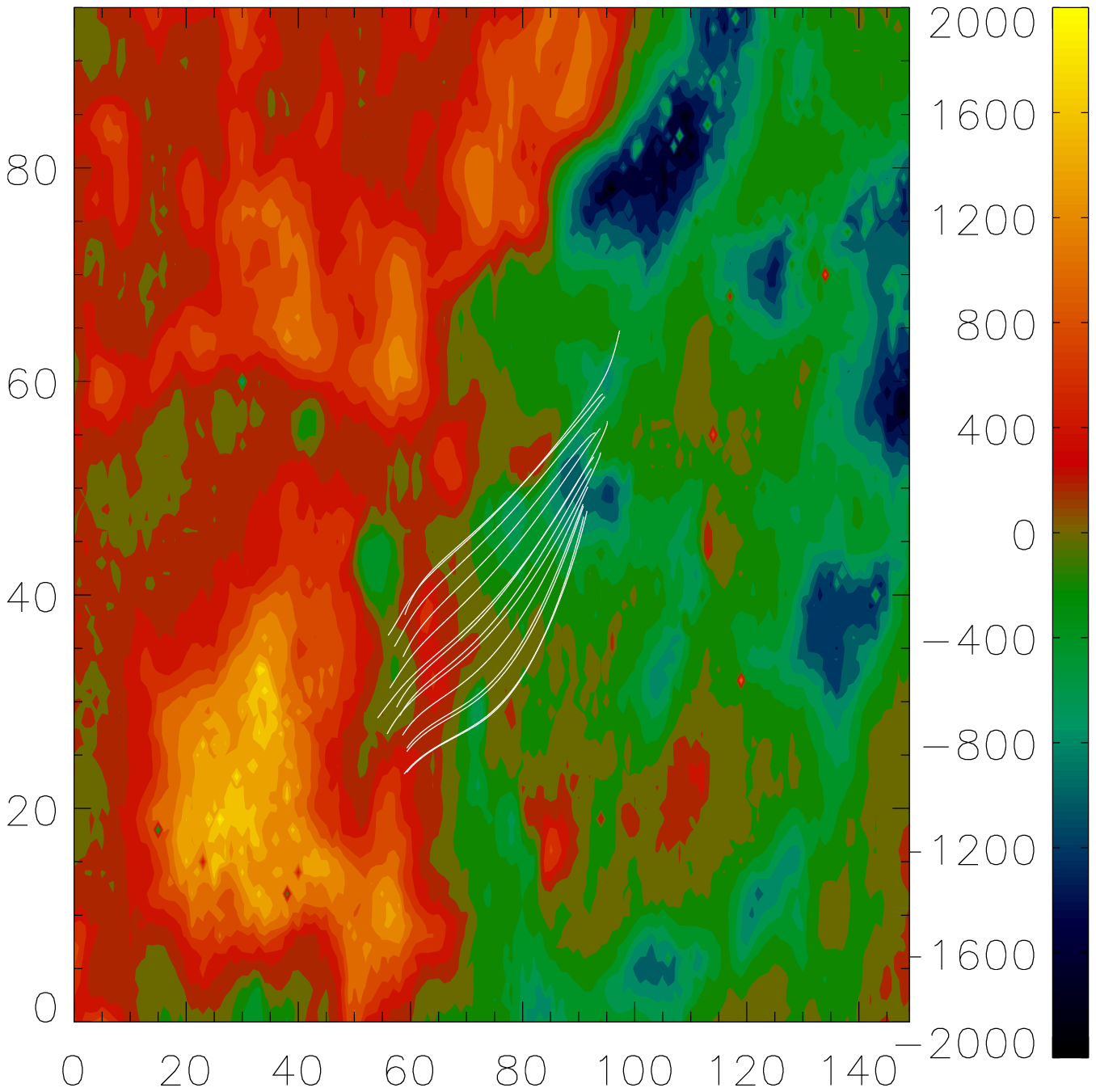}
}
\mbox{
\includegraphics[bb=40 15 380 195,clip,height=4cm]{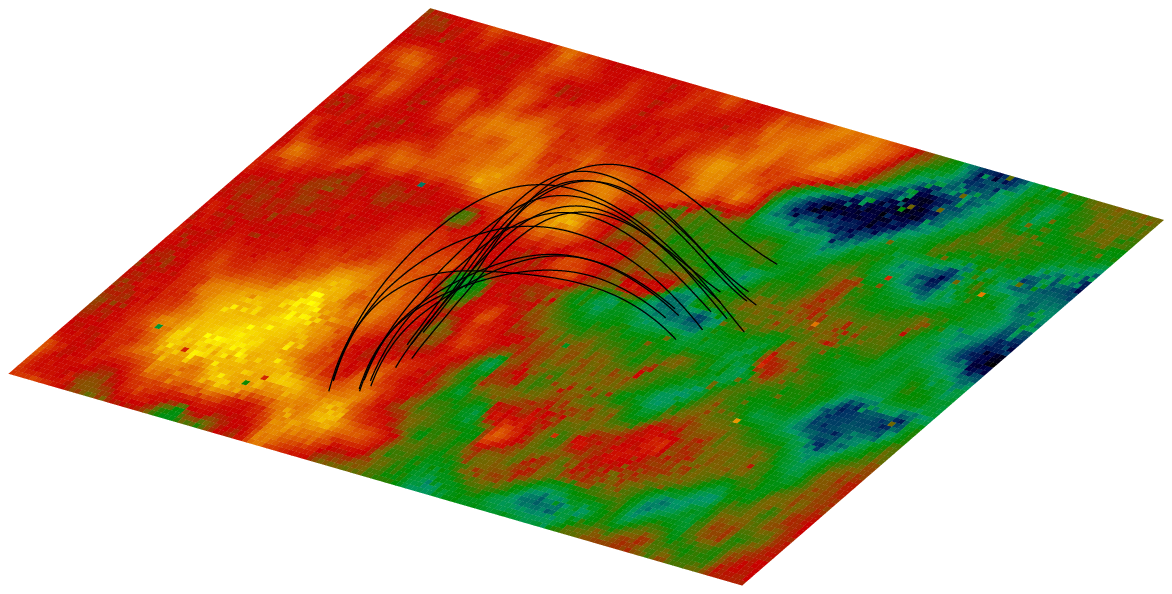}
\includegraphics[bb=120 100 270 290,clip,height=4cm,width=7cm]{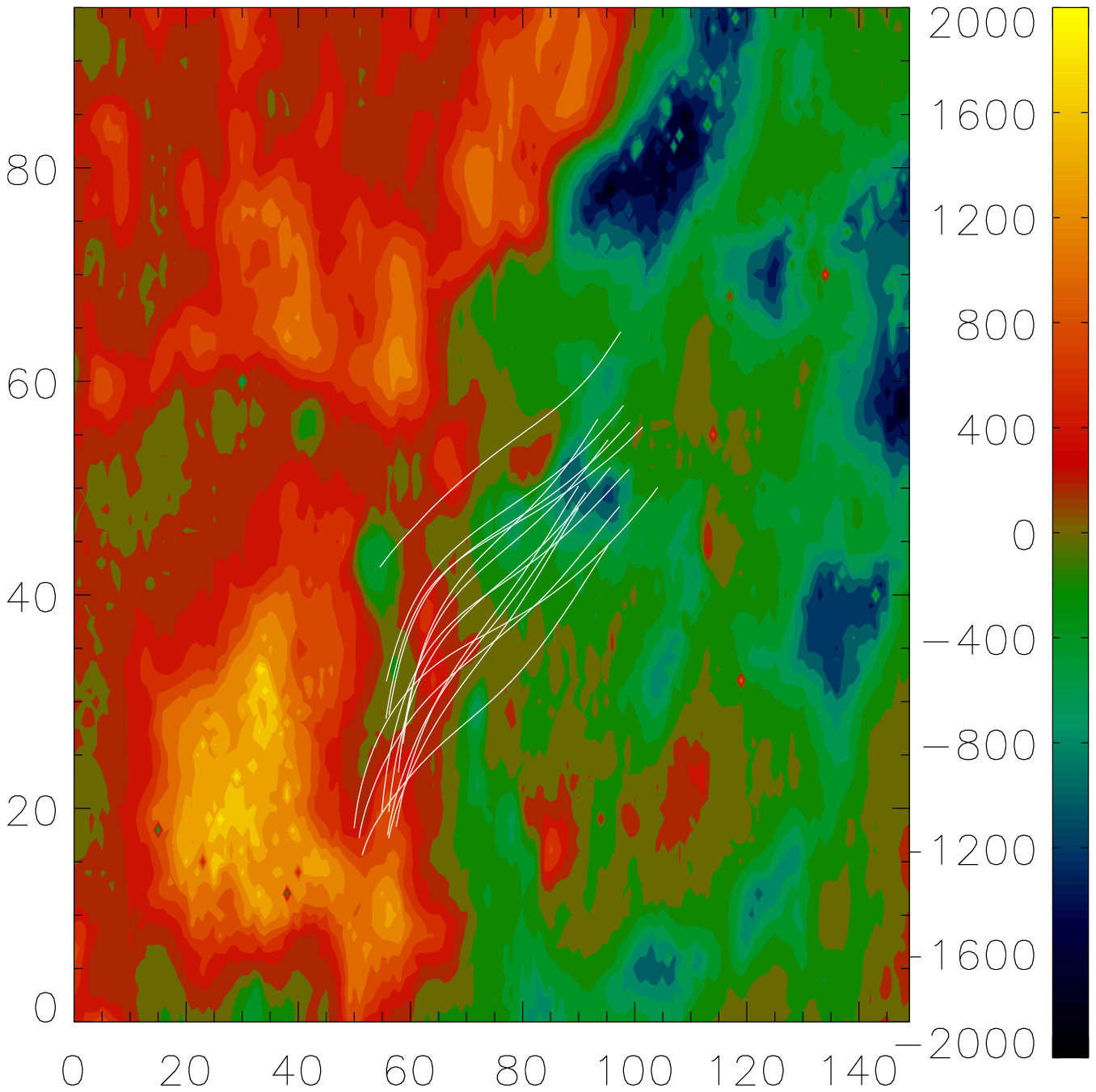}
}
\mbox{
\includegraphics[bb=40 15 380 195,clip,height=4cm]{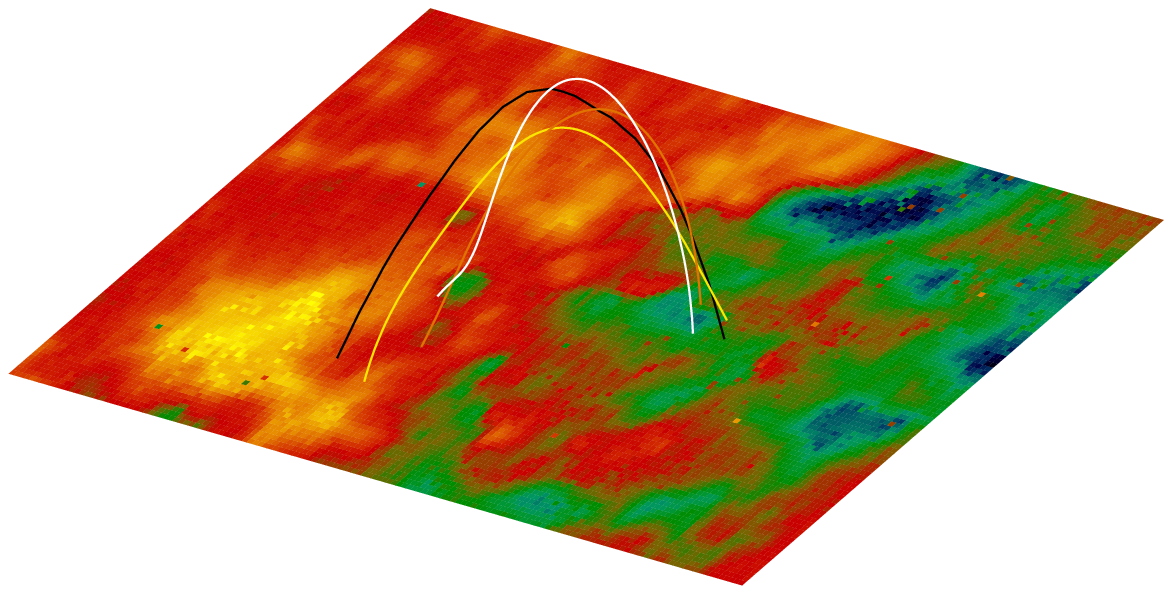}
\includegraphics[bb=120 100 270 290,clip,height=4cm,width=7cm]{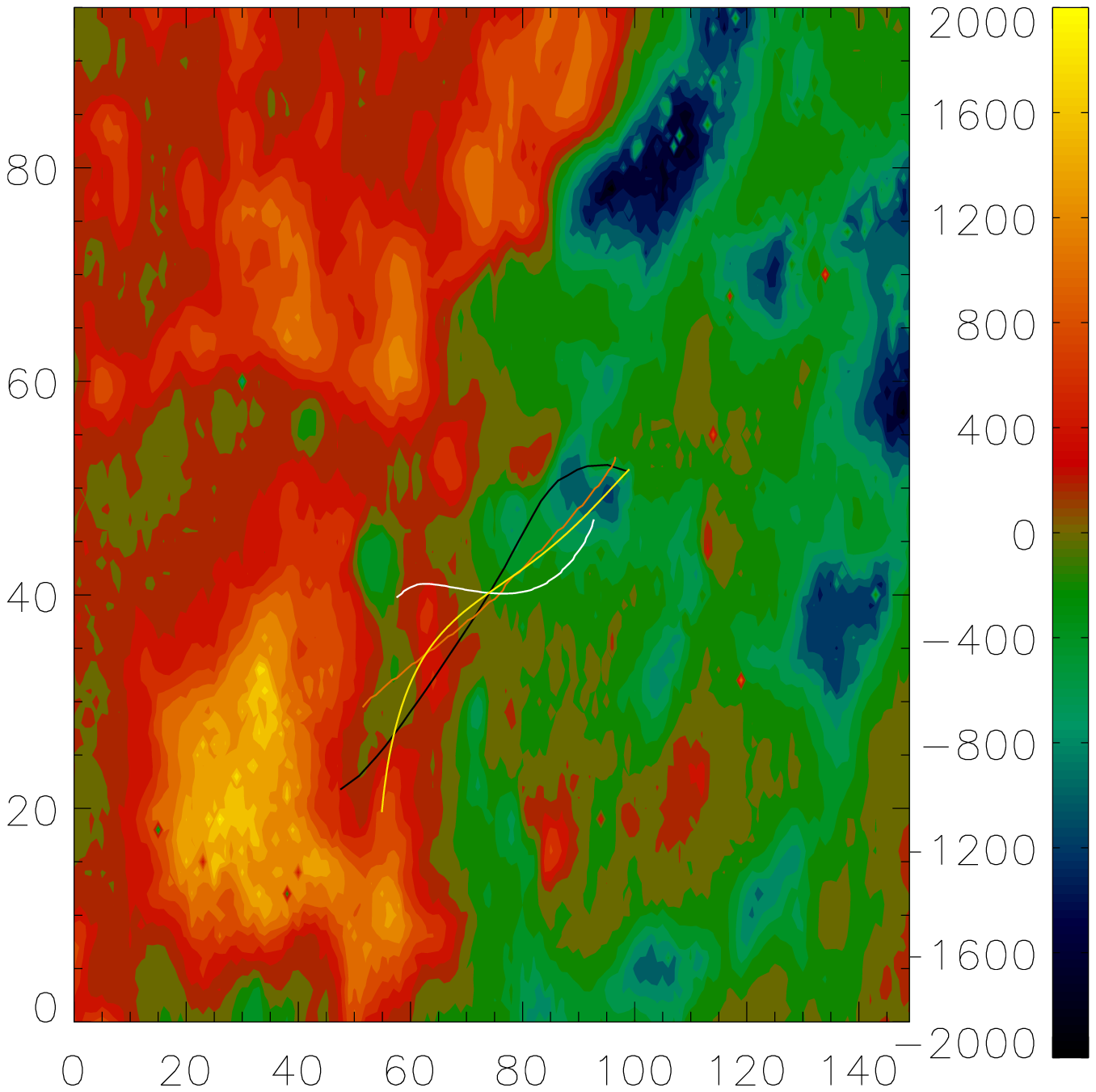}
}
\includegraphics[bb=40 1 360 50,clip,height=1.8cm,width=18cm]{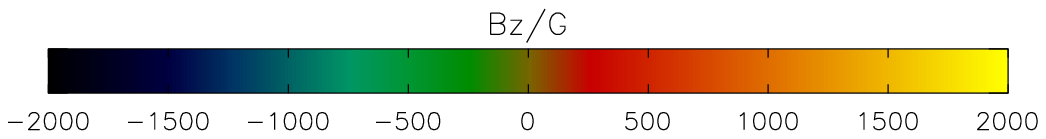}
\caption {Observed and extrapolated fields.
The left panels show the field lines in
3D and the right panels projections on the photosphere. The photospheric LOS field
is represented by colour-coding.
Top panels: Original observed loops.
We show 14 out of a total of 39 loops studied here.
Second row: potential field
extrapolation.  Third row: linear force-free field extrapolation with $\alpha
\cdot L=3.0$.  Fourth: non-linear force-free field extrapolation.
The fifth row shows one loop (Black: original,
white: potential field, orange: linear force-free field with $\alpha \cdot L=3.0$,
yellow: non-linear force-free field.)}
\label{figure1}
\end{figure*}
To compare the reconstructed magnetic field with the observed magnetic loops
we compute magnetic field lines from the reconstructed fields using a
fourth order Runge-Kutta field-line tracer.  The field-line tracer starts the
integration at any arbitrary point in space and traces the magnetic field in
the $+B$ and $-B$ direction until the photosphere is reached in both directions.
As a measure of how well the magnetic field lines and the observed loops agree,
we compute the spatial distance of the two curves in 3D integrated along
the whole loop length $l$ from $\tau=0$ to $\tau=l$. As a result we get a
dimensionless number $C=\frac{1}{l^2} \int_0^{l} \sqrt{\left({\bf
r_{\rm obs}}(\tau)-{\bf r_{\rm extrapol}}(\tau) \right)^2} d \tau$, where
$\tau$ is the geometrical length measured along the loop and $C=0$ if
both curves coincide.  This comparison method has been previously used by
\cite{twtn02} to compare magnetic field lines with stereoscopically observed
loops.

We start the field line integration at 20 points each chosen to lie along an
observed field line ${\bf r_{\rm obs}}(\tau)$ and compute 20 corresponding field lines.
We compare the shapes of these computed field lines (loops) with the observed
field line and
compute the quantitative measure $C$. The lowest value of $C$ corresponds to
the optimal computed magnetic field line.  For potential fields and non-linear
force-free fields the choice of the starting point is the only free parameter
 and finding
the optimal field line is a one-dimensional minimization problem. Linear
force-free fields have the free parameter $\alpha$ and computing the optimal
linear force-free field line is a two-dimensional minimization problem with
respect to $\alpha$ and the integration starting point.
%

\begin{figure}
\includegraphics[width=8cm,height=7cm]{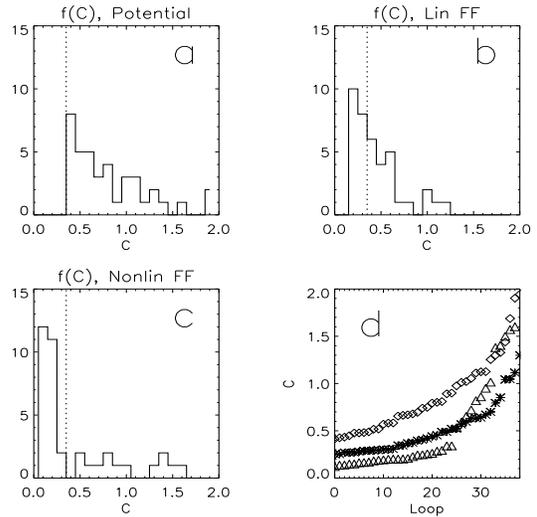}
\caption{
Correspondence of extrapolated loops and measured loops for all
$39$ observed loops. The panels (a)-(c)
show distributions over $C$  for potential, linear and non-linear
force-free fields, respectively.
Panel (d) shows the value for all loops.
The rhombi correspond to a potential field
reconstruction, the stars to the optimal linear force-free reconstruction
and the triangles to a non-linear force-free
reconstruction. (See also Table \ref{tabel1} for the $C$-values of the
individual loops.)}
\label{figureA}
\end{figure}

Fourteen representative loops are shown in Fig. \ref{figure1}
with observed, potential, linear and non-linear force-free loops
being plotted from top to bottom. This figure shows, that force-free
extrapolations are superior to potential fields, with non-linear
force-free being apparently the closest to the observations.
A more quantitative measure is provided by the $C$ values, which
are given for all 39 observed loops in Fig.
\ref{figureA} and Table \ref{tabel1}. The lower the value of
$C$ the better the reconstructed loops
agree with the observed loops. A value of $C \leq 0.35$ seems to be
acceptable.

We find that the simplest magnetic field model, potential fields
(second row in Fig. \ref{figure1}, Fig. \ref{figureA} a),
 provides no agreement with any
observed loop (top row in Fig. \ref{figure1}) within the $C \leq 0.35$
limit.

The inclusion of electric currents, in lowest order with the linear
force-free approximation (Fig. \ref{figureA} b) provides better results, and
for $35 \%$ of the observed loops we get satisfying agreement with the
observations.  One has to keep in mind, however, that a consistent linear
force-free reconstruction requires a unique value of $\alpha$ in the entire
considered volume. Most of the loops have an optimum value of $\alpha L$ in
the range $3-4$ and the quality criterion $C$ only changes slightly within
this range, so that a unique value of $\alpha$ in this range does not give
significantly worse results than the optimal value of $\alpha$. The linear
force-free fields in Fig. \ref{figure1}, third row have been computed with
$\alpha L=3$.

\begin{figure}[h]
\includegraphics[bb=42 42 340 404,clip,width=8cm,height=8cm]{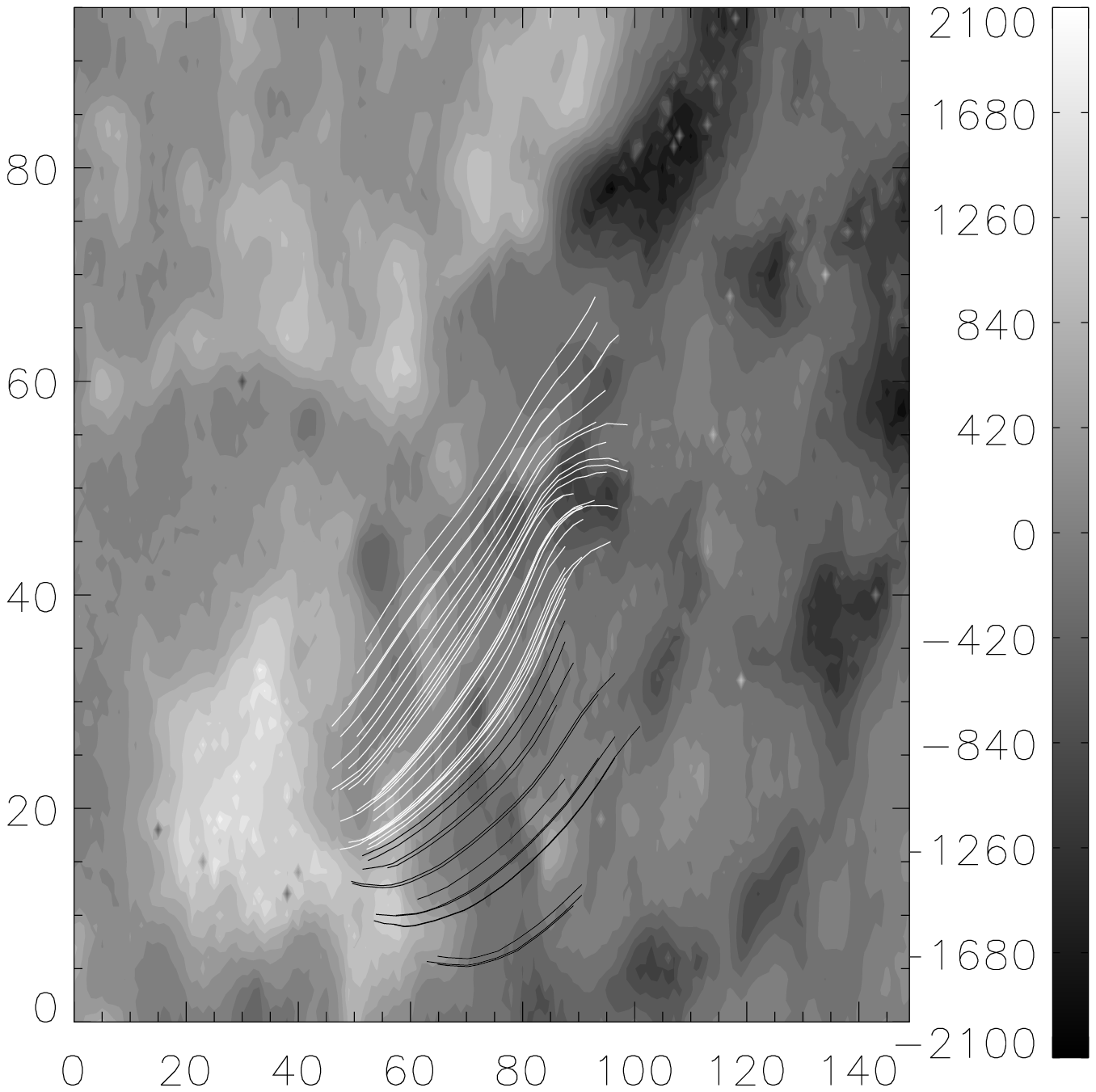}
\caption{
A projection of all observed loops on the photosphere.
We can reconstruct the white loops with the help of a non-linear force-free
field extrapolation with an accuracy of $C < 0.35$.
The black loops cannot be reconstructed with adequate accuracy.}
\label{Appendixfig1}
\end{figure}

The most involved model used here, the non-linear force-free approach
(fourth row in Fig. \ref{figure1} and Fig. \ref{figureA} c), gives
even better results than the linear force free approach. We get a suitable
agreement with the observed loops for $64 \%$ of the loops within the $C \leq
0.35$ limit. Let us  remark that all observed loops which cannot be
reconstructed with this model have at least one foot point close to the
boundary of the available vector magnetic field data.
(see Fig. \ref{Appendixfig1}).

\begin{table} 
\caption{{\small The first column corresponds to an
arbitrary loop number of the observed loops, the second column compares the
measured loops with a potential field extrapolation.  Cols. 3 and 4
compare the observed loops with linear force-free fields and $\alpha L =3$ and
$\alpha L=4$, respectively.  The fifth column contains the optimal linear
force-free parameter $\alpha L$ and the sixth column the comparison with the
observed loops for this optimal value. In the last column we compare the
observations with non-linear force-free reconstructed magnetic loops.}}
\label{tabel1}
\begin{tabular}{|r|r|r|r|r|r|r|}
\hline
Nr& $C_{\rm pot}$ &  $C_{\rm lin}$ &  $C_{\rm lin}$ &
$\alpha_{\rm opt}$ &  $C_{\rm lin}$& $C_{\rm nonlin}$  \\
&&  {\tiny $\alpha L =3$}&{\tiny $\alpha L=4$}&
&{\tiny $\alpha_{\rm opt}$}&   \\
\hline
 0 &$ 1.69 $ &$ 1.40 $ &$ 1.06 $ &$  4.2 $ &$ 1.04 $ &$ 1.60 $\\
  1 &$ 1.13 $ &$ 0.83 $ &$ 0.86 $ &$  2.6 $ &$ 0.79 $ &$ 0.86 $\\
  2 &$ 0.48 $ &$ 0.30 $ &$ 0.28 $ &$  3.5 $ &$ 0.27 $ &$ 0.15 $\\
  3 &$ 0.57 $ &$ 0.34 $ &$ 0.31 $ &$  3.7 $ &$ 0.30 $ &$ 0.14 $\\
  4 &$ 0.75 $ &$ 0.64 $ &$ 0.65 $ &$  3.2 $ &$ 0.63 $ &$ 0.65 $\\
  5 &$ 0.80 $ &$ 0.52 $ &$ 0.45 $ &$  3.8 $ &$ 0.45 $ &$ 0.19 $\\
  6 &$ 0.69 $ &$ 0.37 $ &$ 0.31 $ &$  4.1 $ &$ 0.30 $ &$ 0.17 $\\
  7 &$ 0.90 $ &$ 0.56 $ &$ 0.43 $ &$  4.3 $ &$ 0.40 $ &$ 0.12 $\\
  8 &$ 0.58 $ &$ 0.35 $ &$ 0.30 $ &$  3.8 $ &$ 0.29 $ &$ 0.14 $\\
  9 &$ 1.90 $ &$ 1.57 $ &$ 1.39 $ &$  4.2 $ &$ 1.05 $ &$ 1.38 $\\
 10 &$ 0.68 $ &$ 0.44 $ &$ 0.51 $ &$  3.0 $ &$ 0.44 $ &$ 0.59 $\\
 11 &$ 0.43 $ &$ 0.26 $ &$ 0.29 $ &$  2.9 $ &$ 0.25 $ &$ 0.25 $\\
 12 &$ 0.98 $ &$ 0.56 $ &$ 0.73 $ &$  3.5 $ &$ 0.52 $ &$ 1.01 $\\
 13 &$ 0.89 $ &$ 0.86 $ &$ 0.87 $ &$  3.5 $ &$ 0.85 $ &$ 0.95 $\\
 14 &$ 1.12 $ &$ 0.66 $ &$ 0.53 $ &$  3.7 $ &$ 0.52 $ &$ 0.26 $\\
 15 &$ 1.05 $ &$ 0.63 $ &$ 0.49 $ &$  4.0 $ &$ 0.49 $ &$ 0.20 $\\
 16 &$ 0.66 $ &$ 0.35 $ &$ 0.29 $ &$  4.1 $ &$ 0.29 $ &$ 0.15 $\\
 17 &$ 0.43 $ &$ 0.27 $ &$ 0.28 $ &$  3.0 $ &$ 0.27 $ &$ 0.21 $\\
 18 &$ 0.41 $ &$ 0.26 $ &$ 0.26 $ &$  3.3 $ &$ 0.25 $ &$ 0.16 $\\
 19 &$ 0.48 $ &$ 0.30 $ &$ 0.30 $ &$  3.3 $ &$ 0.28 $ &$ 0.16 $\\
 20 &$ 0.67 $ &$ 0.35 $ &$ 0.31 $ &$  4.4 $ &$ 0.29 $ &$ 0.19 $\\
 21 &$ 0.49 $ &$ 0.34 $ &$ 0.30 $ &$  3.6 $ &$ 0.29 $ &$ 0.20 $\\
 22 &$ 0.74 $ &$ 0.53 $ &$ 0.42 $ &$  4.4 $ &$ 0.42 $ &$ 0.18 $\\
 23 &$ 0.45 $ &$ 0.31 $ &$ 0.28 $ &$  3.3 $ &$ 0.27 $ &$ 0.34 $\\
 24 &$ 0.48 $ &$ 0.35 $ &$ 0.31 $ &$  3.4 $ &$ 0.31 $ &$ 0.29 $\\
 25 &$ 0.52 $ &$ 0.38 $ &$ 0.36 $ &$  4.2 $ &$ 0.34 $ &$ 0.20 $\\
 26 &$ 0.52 $ &$ 0.38 $ &$ 0.34 $ &$  4.0 $ &$ 0.34 $ &$ 0.26 $\\
 27 &$ 0.58 $ &$ 0.40 $ &$ 0.48 $ &$  2.6 $ &$ 0.37 $ &$ 0.24 $\\
 28 &$ 0.66 $ &$ 0.45 $ &$ 0.57 $ &$  2.5 $ &$ 0.39 $ &$ 0.23 $\\
 29 &$ 0.81 $ &$ 0.61 $ &$ 0.84 $ &$  2.8 $ &$ 0.60 $ &$ 0.82 $\\
 30 &$ 1.01 $ &$ 0.68 $ &$ 1.25 $ &$  3.2 $ &$ 0.67 $ &$ 1.40 $\\
 31 &$ 1.44 $ &$ 0.98 $ &$ 0.80 $ &$  3.6 $ &$ 0.63 $ &$ 2.73 $\\
 32 &$ 1.33 $ &$ 0.70 $ &$ 0.79 $ &$  3.0 $ &$ 0.70 $ &$ 0.72 $\\
 33 &$ 1.30 $ &$ 0.71 $ &$ 0.71 $ &$  3.2 $ &$ 0.64 $ &$ 0.53 $\\
 34 &$ 1.25 $ &$ 0.72 $ &$ 0.60 $ &$  3.4 $ &$ 0.57 $ &$ 0.34 $\\
 35 &$ 1.02 $ &$ 0.64 $ &$ 0.49 $ &$  4.0 $ &$ 0.49 $ &$ 0.20 $\\
 36 &$ 1.94 $ &$ 1.66 $ &$ 1.32 $ &$  4.2 $ &$ 1.30 $ &$ 1.57 $\\
 37 &$ 0.79 $ &$ 0.48 $ &$ 0.37 $ &$  4.0 $ &$ 0.37 $ &$ 0.28 $\\
 38 &$ 1.11 $ &$ 1.27 $ &$ 1.30 $ &$ -0.2 $ &$ 1.11 $ &$ 1.50 $\\
\hline
\end{tabular}
\end{table}

\section{Conclusions}
\label{results}
We have compared the observationally inferred structure of magnetic loops in
the upper chromosphere with magnetic fields extrapolated from photospheric
measurements.  We find that the simplest model, potential fields, is not
sufficient to reproduce the observations.  The inclusion of field line
parallel currents, the so called force-free approach, gives much better
results. Among force-free models, a linear one gives less accurate results
than the non-linear force-free extrapolation.

We find that the observed and extrapolated loops
agree quite well for almost $2/3$ of the loops, while the remaining $1/3$
might suffer from the limited field of view of the available vector
magnetogram.

The investigated active region is quite young. With a vertical upflow speed
of $v=1.5{^+_-}0.5 {\rm km/s}$ at the loop apex and a maximum loop height
of $10 {\rm Mm}$ we can estimate the time elapsed since the loop tops first
emerged as $2 {\rm h}{^+_-}40 {\rm min}$ assuming a constant rise speed.
A horizontal shear flow of $1 {\rm km/s}$ on the photosphere would give a shear of
$7.2 {\rm Mm}$.
This value is comparable to the difference of the footpoint
locations between potential field loops and observed loops.
It is therefore not clear, whether the electric current has been
caused by shear flow motion or if the magnetic loops already contain the current
during their emergence.
%
The rise of the loops may also explain some of the discrepancy between
observed and extrapolated loops, since the loops may change somewhat
during the time needed for the instrument to scan the region.
\begin{acknowledgements}
The work of Wiegelmann was supported by  DLR-grant 50 OC 0007.
We thank an unknown referee for useful remarks.
\end{acknowledgements}

\clearpage

\end{document}